# High-Q Millimeter-Wave Acoustic Resonators in Thin-Film Lithium Niobate Using Higher-Order Antisymmetric Modes

Vakhtang Chulukhadze, Jack Kramer, Tzu-Hsuan Hsu, Omar Barrera, and Ruochen Lu

***Abstract*—This letter presents miniature millimeter-wave (mmWave, above 30 GHz) acoustic resonators based on a thin-film lithium niobate (LN) platform. More specifically, we present high performance third-order antisymmetric (A3) mode laterally excited bulk acoustic wave resonators (XBAR) from 30 to 50 GHz. Compared to prior demonstrations, the proposed platform features a compact footprint owing to a smaller lateral wavelength and aperture. We showcase an A3 mode device operating at 39.8 GHz with a high extracted electromechanical coupling ($k^2$) of 4%, a high 3-dB series resonance quality factor ($Q_s$) of 97, and a high 3-dB anti-resonance quality factor ($Q_p$) of 342, leading to a figure of merit (FoM=$Q{\cdot}k^2$) of 13.8 with a compact footprint of 32x44 μm². To demonstrate frequency scalability, the piezoelectric film thickness is varied while keeping the device layout. As a result, we present a multitude of miniature, high-performance devices covering a wide frequency span of 30-50 GHz, validating the proposed XBAR design at mmWave.**

## I. Introduction

WITH the pervasiveness of 5G wireless communication, compact front-end signal processing components at millimeter-wave (mmWave, above 30 GHz) have garnered attention [1]–[4]. Below 6 GHz, surface acoustic wave (SAW) and bulk acoustic wave (BAW) devices - based on piezoelectric materials such as aluminum nitride (AlN), scandium aluminum nitride (ScAlN), lithium niobate (LN), and lithium tantalate (LT) - have been instrumental in realizing miniature filters for mobile applications [5]–[7]. However, scaling these technologies to mmWave introduces intrinsic limitations, posing a challenge for the next generation of high-frequency mobile communication systems [8], [9].

Despite significant advances in state-of-the-art radio frequency acoustic technologies, inherent limitations to frequency scaling continue to motivate investigations into alternative platforms. Solidly mounted devices have recently achieved very high phase velocities, but they necessitate sub-100 nm feature patterning alongside expensive substrates, e.g. silicon carbide (SiC) and diamond [10], [11]. Cross-sectional Láme mode resonators exhibit high phase velocity and $Q$, yet they also require fine features [12]–[14]. Film bulk acoustic wave resonators (FBAR) exhibit potential for mmWave utilizing acoustic overtones or periodically poled piezoelectric films (P3F), but they face challenges with moderate $Q$, primarily due to the large metal to piezoelectric volume ratio at the required film thicknesses [15].

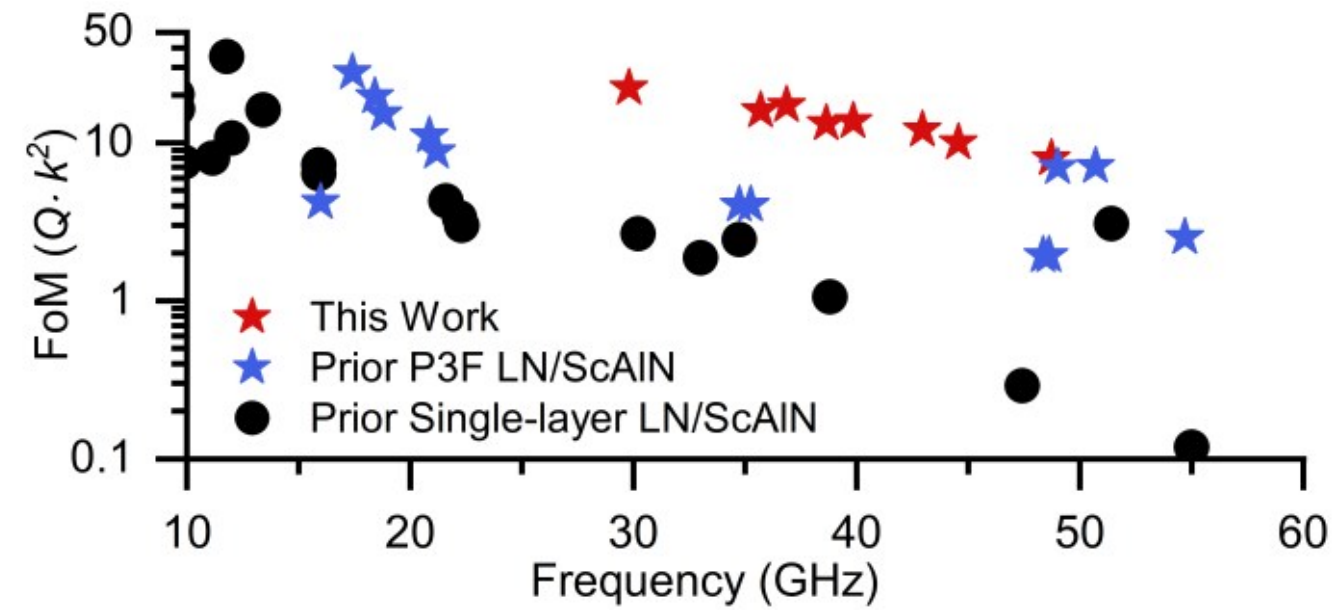

Fig. 1. A survey of acoustic resonator performance at 10-60 GHz, including prior single-layer LN [16]–[19], single-layer AlN and ScAlN [13]–[15], [20], [21], P3F LN [22]–[25], and P3F ScAlN [26]–[28]. We use the obtained 3-dB $Q_p$ to compare with previously reported FoMs, where the higher value between the self-reported $Q_s$ and $Q_p$ was used.

Alternatively, recent progress in thin-film transfer technology has enabled first-order antisymmetric (A1) mode laterally excited acoustic wave resonators (XBAR) in lithium niobate (LN), without the need for a bottom electrode [4]. However, while recent mmWave XBAR prototypes have reported high electro-mechanical coupling ($k^2$), they have shown moderate $Q$ [16]–[19], yielding a moderate figure of merit (FoM, $Q{\cdot}k^2$, Fig. 1). To address the limitations, recent efforts have demonstrated P3F LN XBARs, achieving $Q$ as high as 237 at 50 GHz using bilayer LN [22]–[25]. Nevertheless, the P3F structure introduces new fabrication limitations, particularly in maintaining precise thickness control over each layer. These implications motivate a revisit of single-layer XBARs with a new design strategy.

In this work, we present high-performance single-layer mmWave acoustic resonators by minimizing their footprint and maximizing the device volume-to-area ratio. We present devices spanning 30-50 GHz, covering a broad portion of the mmWave spectrum. Notably, we highlight a miniature XBAR at 39.83 GHz with a 3-dB series resonance ($f_s$) quality factor ($Q_s$) of 97, a high 3-dB antiresonance ($f_p$) quality factor ($Q_p$) of 342, and an extracted $k^2$ of 4%. Based on the extracted $k^2$ and

The paper was submitted on 07/21/2025. This work was supported by the DARPA COFFEE project.

All authors are with the University of Texas at Austin Department of Electrical and Computer Engineering

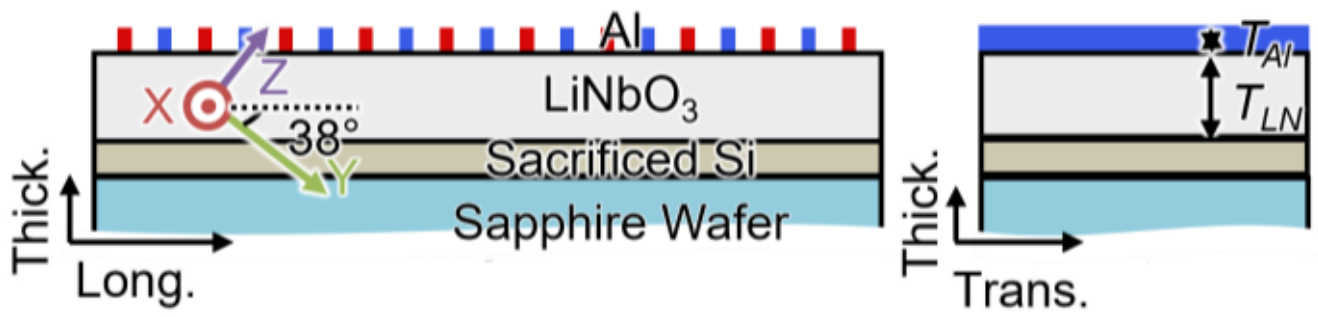

Fig. 2 The cross-sectional view of the chosen acoustic stack.

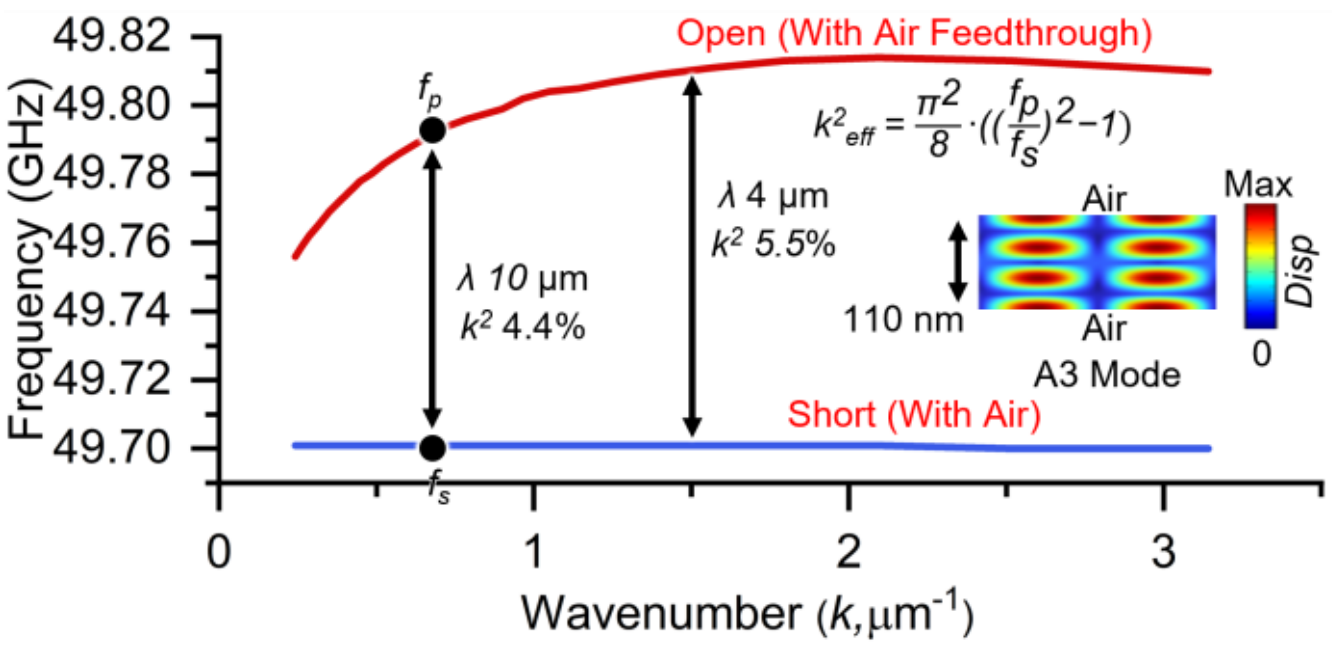

Fig. 3. A3 Mode dispersion analysis in 128Y LN under electrically open and short boundary conditions with air feedthrough. Calculated coupling is labeled.

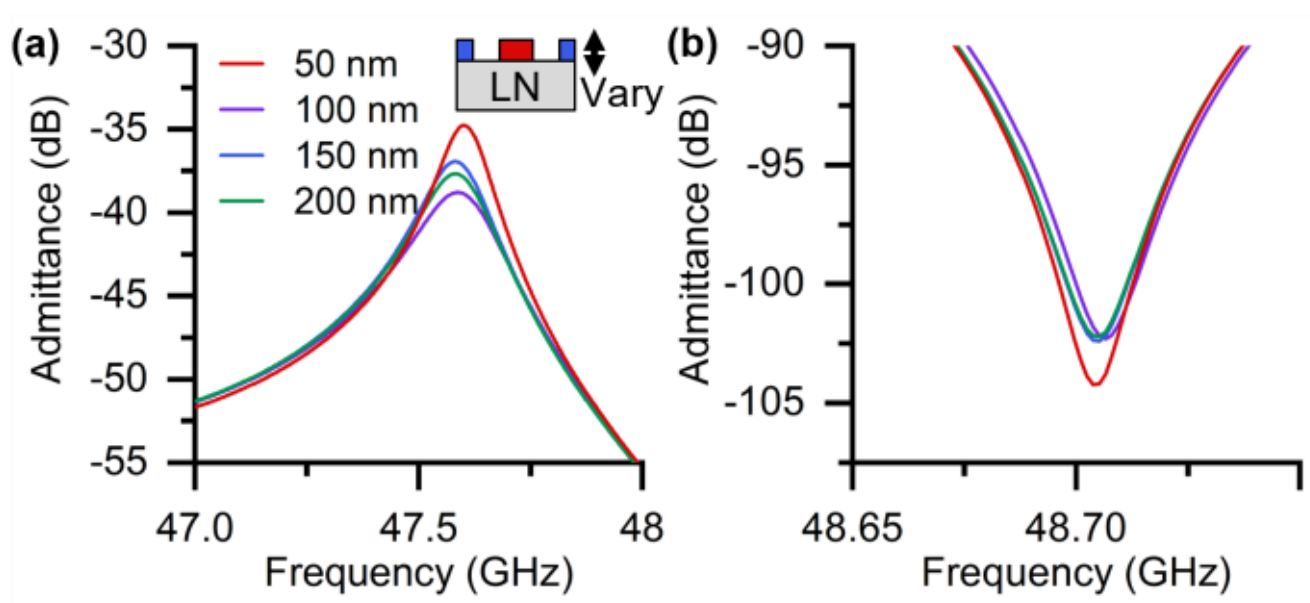

Fig. 4. FEA simulations of resonators with varying electrode thicknesses at (a) series and (b) parallel resonances, indicating the effects of thin electrodes.

3-dB $Q_p$, the device achieves a high FoM of 13.8 with a compact footprint of 32×44 µm$^2$.

## II. Device Design

Suspended acoustic resonators were designed in a 128Y LN/amorphous silicon/sapphire stack (Fig. 2). We selected a sample with LN film thickness ranging from 110 to 160 nm to investigate device performance from 30 to 50 GHz.

Diverging from the conventional XBAR design space with long lateral wavelengths ($\lambda_{Lateral}$), the devices were made compact by reducing $\lambda_{Lateral}$ to 3.15 µm and 4 µm. This choice was made by inspecting the third-order antisymmetric Lamb mode (A3) dispersion curves, enabling a trade-off between $\lambda_{Lateral}$ to achieve a higher $k^2$ (Fig. 3). While prior research indicates that a higher $\lambda_{Lateral}$ is preferred due to superior spurious-mode performance for A1, we specifically tailored the design for A3 operation at mmWave [29]. Utilizing the low chosen $\lambda_{Lateral}$, the device footprint was reduced for compactness and to minimize surface losses in the acoustic resonator, which arise from both the electrical and the mechanical domains and significantly impact overall device performance [30].

Next, the electrode material and thickness were examined to account for their substantial impact on device performance at reduced piezoelectric film thicknesses. Aluminum (Al) was selected for its excellent conductivity and low mass-loading. The electrode thickness was analyzed using COMSOL finite element analysis (FEA). Strong mechanical damping was assigned in the electrodes, where a low $Q$ of 5 was chosen as a nominal value for the metal to highlight its effects on the overall performance. On the other hand, LN was modelled as a low-loss structure. With this approach, we gained insights into relative trends in $Q$ as the ratio of metal to LN decreased. Simulations conducted with a 50% electrode duty cycle reveal that reducing the electrode thickness yields a higher impedance ratio and consequently leads to a higher overall $Q$ [Fig. 4 (a)(b)]. In this work, we selected an electrode thickness of 50 nm Al. While variations in the duty cycle can further influence impedance characteristics, a 50% duty cycle was maintained while considering its effect on the feature size. Additionally, a minimized device aperture of 8 $\lambda_{Lateral}$ was selected to mitigate the routing resistance from the electrodes. These design choices enhance overall $Q$ while maintaining a balance between performance at $f_s$ and $f_p$.

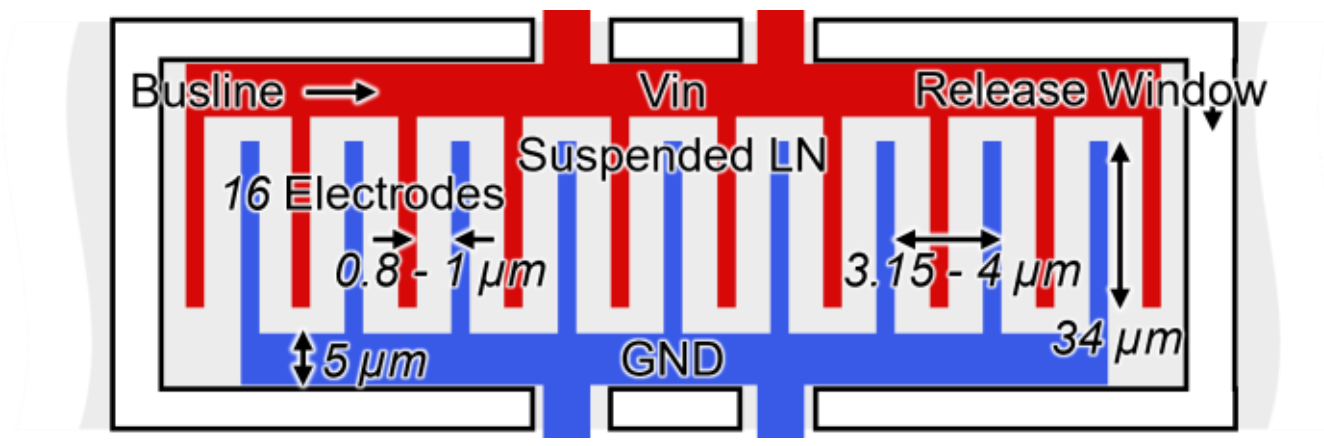

Fig. 5. (a) Top view of the finalized device design. The key design parameters are labeled.

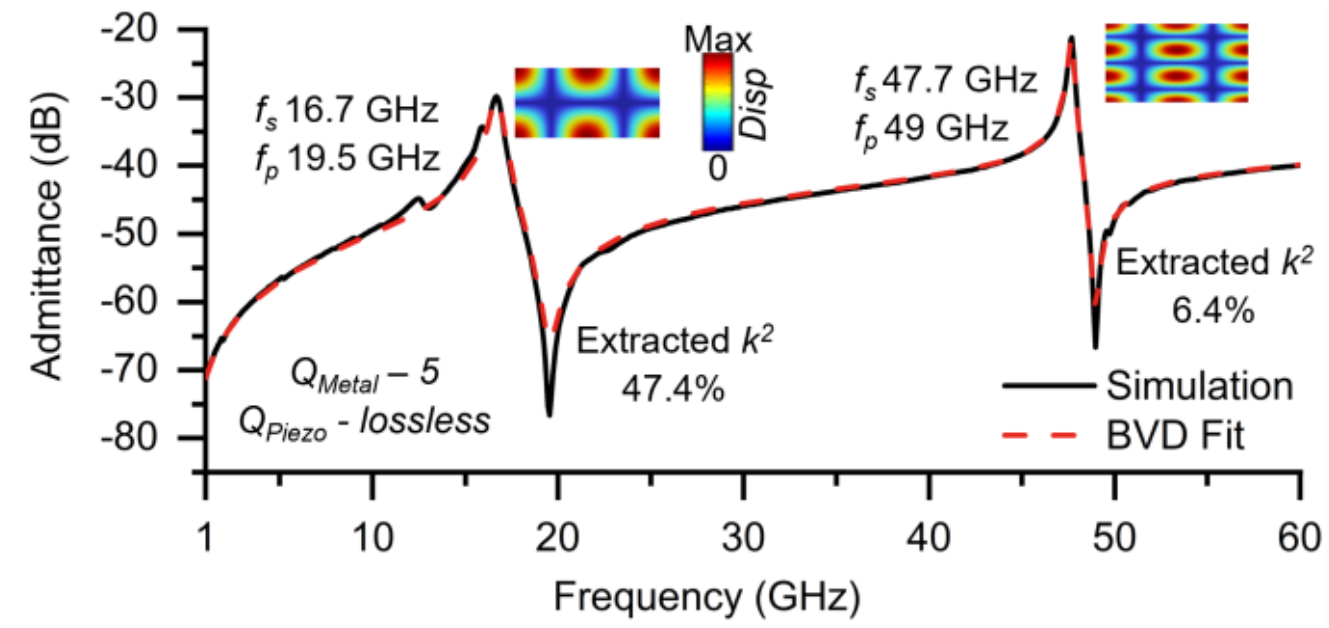

Fig. 6. FEA simulation using the parameters listed in Fig. 5.

The top view of the resulting design can be seen in Fig. 5, where key design parameters are labeled. The FEA simulated admittance with a 4 µm $\lambda_{Lateral}$ shows mmWave operation and a high extracted $k^2$ of 6.4% at 47.7 GHz (Fig. 6), which is obtained from an equivalent Butterworth-Van-Dyke (BVD) model with two motional branches.

## III. Fabrication and Experimental Results

The fabrication steps are outlined in Fig. 7 (a). The process follows closely with [31]. A scanning electron microscope image is in Fig. 7 (b). The XBAR resonators were measured using a Keysight vector network analyzer (VNA) at an input power of −15 dBm, and calibrated with a GGB CS-5 substrate. Due to the miniature footprint, the devices exhibited a parasitic electromagnetic resonance well above their acoustic response. Accordingly, we report raw data from the calibrated VNA.

The measured admittance magnitude and phase of a representative device are in Fig. 8. The device exhibited $f_s$ at 39.83 GHz with a 3-dB $Q_s$ of 97 and $f_p$ at 40.67 GHz with a 3-dB $Q_p$ of 342. The measurement is further fit with an mBVD model with series routing resistance ($R_s$) and inductance ($L_s$)

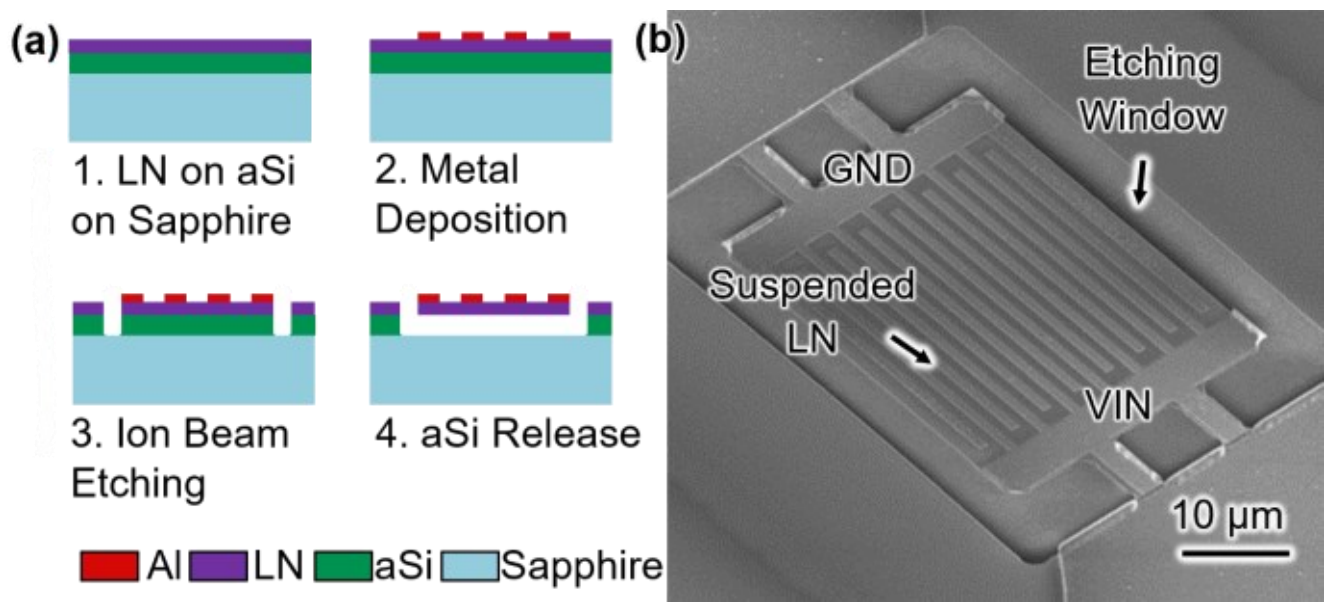

Fig. 7 (a) The fabrication flow for this work. (b) A Scanning electron microscope (SEM) image of the fabricated device.

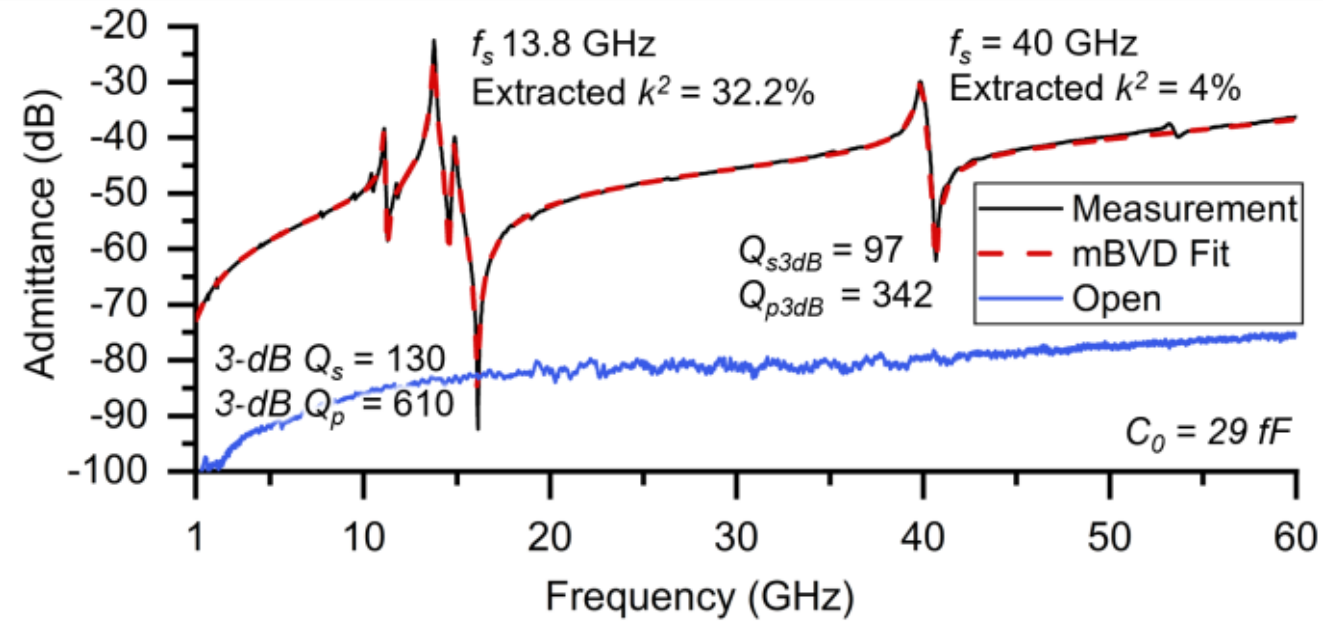

Fig. 8. Measured wide-band response alongside an mBVD fit accounting for routing resistance and inductance. Note that the high $Q_p$ for A1 mode is not realistic, as the impedance is too high for VNA testing. On the other hand, the $Q_p$ for A3 is within the VNA dynamic range.

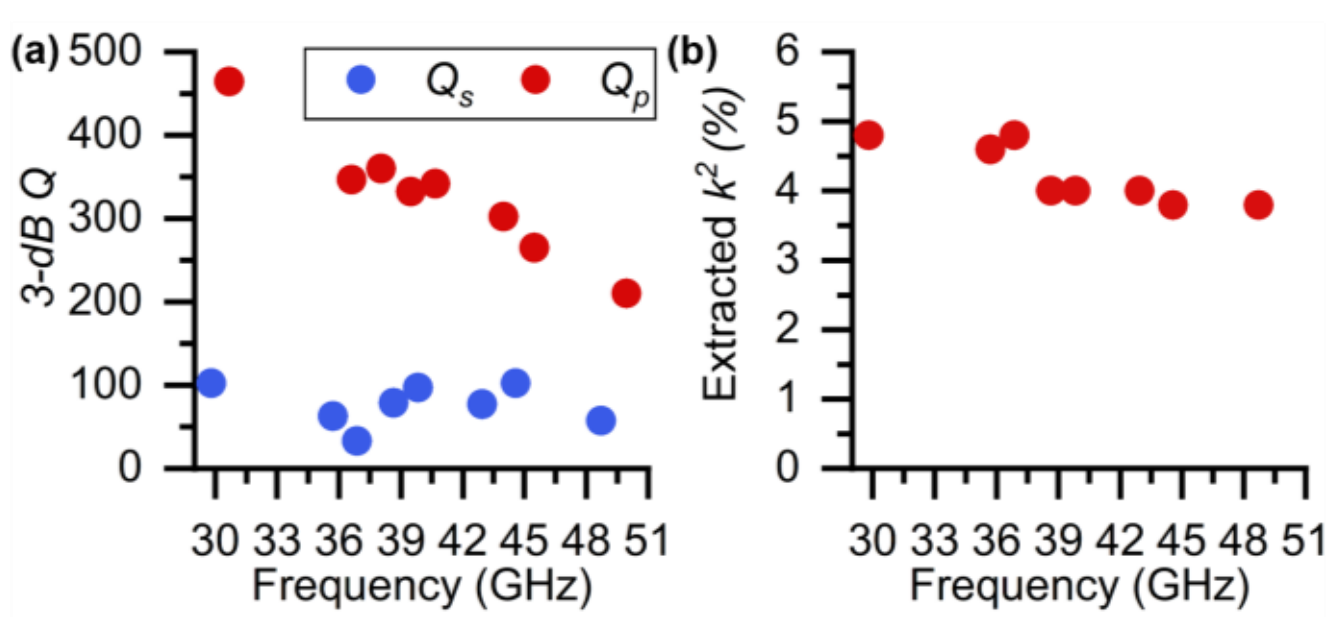

Fig. 9. (a) Measured 3-dB $Q_s$ and $Q_p$, indicating high performance from 35-50 GHz. (b) Extracted $k^2$ from mBVD.

[25]. The mBVD fit shows good agreement with the measured data. The A3 XBAR showed a motional inductance ($L_m$) of 19 nH, a motional capacitance ($C_m$) of 0.8 fF, a motional resistance ($R_m$) of 18.6 Ω, series resistance ($R_s$) of 18.42 Ω, a series inductance ($L_s$) of 0.1 nH, mBVD $Q$ ($Q_{mBVD}$) of 260, and an extracted $k^2$ of 4%. $Q_p$ and $Q_{mBVD}$ are slightly different, due to the limited motional branches in the fit. In comparison with the SoA (Fig. 1), the high achieved 3-dB $Q_p$, alongside the extracted $k^2$, leads to a high FoM of 13.8 at 40 GHz, highlighting the effectiveness of the proposed design. To validate the extracted $Q_p$ of A3 against potential dynamic range limitations of the VNA, we plot the open response (with probe unlanded), which confirms sufficient dynamic range margin for A3, and supports the reliability of the $Q_p$ extraction.

The higher measured $Q_s$ of A1 over A3 suggests increased resistive, EM, and mechanical losses at higher frequencies. To clarify, Fig. 4 only considers the mechanical damping, indicating that mechanical loss in the electrodes affects $Q_s$ disproportionally compared with $Q_p$, which is a previously underappreciated loss mechanism at series resonance. Still, isolating a single dominant loss mechanism at mmWave remains difficult due to the combined influence of resistive, EM, and surface effects [22], [30], [32]. Addressing this interplay will be crucial to future high-$Q$ resonator designs. Additionally, it is notable that the extracted $k^2$ here is significantly smaller than the perceived $k^2$ directly calculated from $\pi^2/8\cdot((f_p/f_s)^2-1)$, due to the EM self-resonance [25]. Accordingly, we report only the extracted values from fitting to avoid unnecessary overclaims.

Table 1: Comparison to Prior Single-Layer LN XBARs

| Ref. | Lateral Wave. (μm) | Electrode Thick. (nm) | Normalized Footprint (μm²)* | $f_s$ (GHz) | $k^{2**}$ | 3-dB $Q_s$ | 3-dB $Q_p$ | $f\cdot Q^{***}$ |
|---|---|---|---|---|---|---|---|---|
| [33] | 4 | 50 | 3162 | 4 | 28% | 400 | 100 | 1.6E12 |
| [16] | 8 | 350 | 1377 | 57 | 5.5% | 60 | 60 | 3.4E12 |
| [17] | 20 | 50 | 1920 | 12.9 | 3.8% | 282 | 214 | 3.7E12 |
| [18] | 13 | 60 | 2028 | 12.8 | 3.7% | 224 | 224 | 2.9E12 |
| [19] | 34 | 310 | 36000 | 33.7 | 3.7% | 67 | 67 | 2.3E12 |
| **This Work** | **4** | **50** | **857** | **29.8** | **4.8%** | **102** | **464** | **1.4E13** |
| | **3.2** | **50** | **626** | **36.8** | **4.8%** | **33** | **360** | **1.3E13** |
| | **4** | **50** | **684** | **40** | **4%** | **96** | **342** | **1.4E13** |
| | **3.2** | **50** | **564** | **48.8** | **3.8%** | **58** | **210** | **1.0E13** |

*Impedance at series resonance normalized to 50 Ω. **$k^2$ is the extracted value from fitting. ***Q is selected as the maximum between reported $Q_s$ and $Q_p$.*

Measurements with the same device layout across 30-50 GHz consistently show high performance (Fig. 9). Across this frequency span, $Q_s$ ranged from 33 to 102, and $Q_p$ from 210 to 464. $Q_p$ decreased linearly with frequency, attributed to the increasing acoustic and dielectric loss at higher frequencies. Conversely, $Q_s$ was predominantly limited by the resistive and mechanical loss arising from the thin Al electrodes. Hence, based on the extracted $k^2$ and 3-dB $Q_p$, we obtain FoMs ranging from 8 to 22.3 within 30-50 GHz (Fig. 1).

Table 1 benchmarks the proposed mmWave resonators against prior single-layer LN XBARs [16]–[19], [33]. Our devices significantly exceed state-of-the-art XBARs in terms of the $f\cdot Q$ product, achieving up to $1.4\times10^{13}$ while maintaining a compact normalized footprint to achieve 50 Ω at $f_s$. These improvements primarily result from key innovations: a compact resonator design thanks to a reduced $\lambda_{Lateral}$ combined with a minimized transverse aperture; and thin (50 nm) electrodes to lower mechanical loss. In tandem, these design enhancements allow resonators to achieve a high $Q_p$ at mmWave frequencies.

Future mmWave LN XBAR iterations will focus on systematically optimizing key device parameters, including electrode thickness, metal composition and quality, the electrode duty cycle, and the anchor geometry. Additionally, we will explore post-processing techniques such as annealing for further potential reduction in loss. Future device scaling for practical filter applications will focus on shorter aperture lengths and optimized electrode arrays, accompanied by thicker buslines to minimize routing resistance and improve $Q$.

## IV. Conclusion

In this work, we have demonstrated miniature, high-performance A3 XBARs from 30 to 50 GHz. The compact structure outperforms the SoA, highlighting the frequency scalability. This alternative approach tailors the device design to achieve high $Q$ at mmWave. With continued optimization, the A3 mode XBAR platform is poised to introduce low-loss acoustic resonators at mmWave frequencies.